\documentclass[prl, reprint]{revtex4-2}
\usepackage{graphicx}
\usepackage{dcolumn}
\usepackage{bm}
\usepackage{xcolor}
\usepackage{placeins}
\usepackage{setspace}
\usepackage{bbold}

\newcommand{\be}{\begin{equation}}
\newcommand{\ee}{\end{equation}}
\newcommand{\bfig}{\begin{figure}}
\newcommand{\efig}{\end{figure}}

\usepackage{lineno}
\usepackage{lipsum}

\begin{document}

\title{The vortex-Nernst effect in a superconducting infinite-layer nickelate}

\author{Nicholas P. Quirk$^{1}$}
\author{Danfeng Li$^{2,3}$}
\author{Bai Yang Wang$^{2,4}$}
\author{Harold Y. Hwang$^{2,3}$}
\author{N. P. Ong$^{1,\S}$}
\affiliation{
{$^1$Department of Physics, Princeton University, Princeton, NJ 08544, USA}\\
{$^2$Stanford Institute for Materials and Energy Sciences, SLAC National Accelerator Laboratory, Menlo Park, California 94025, USA}\\
{$^3$Department of Applied Physics, Stanford University, Stanford, CA 94305, USA}\\
{$^4$Department of Physics, Stanford University, Stanford, CA 94305, USA}\\
}

\date{\today}

\begin{abstract}
We report measurements of the Nernst and Seebeck effects in Nd$_{1-x}$Sr$_x$NiO$_2$ thin films near the superconducting transition temperature, $T$ = 6.5 - 15 K. Our main result is the observation of a vortex-Nernst signal $S_{yx}(T,H)$ with a maximum at $\mu_0H$ = 5 T and a tail that extends to $\mu_0H$ $\approx$ 15 T, which we identify as the upper-critical field $H_{c2}$. At $T > T_c =$ 6.1 K, $H_{c2}$ remains large (15 T), up to the highest temperature we can resolve from $S_{yx}$ (11 K). These results indicate the existence of a vortex-liquid state over a wide range of finite-resistance temperatures, as in the high-$T_{\rm c}$ cuprates.
\end{abstract}

\maketitle
\section{Introduction}
The discovery of superconductivity in hole-doped infinite-layer nickelates has drawn many comparisons to the high-$T_{\rm c}$ cuprate superconductors~\cite{DLi1, KWLee, Botana, Sakakibara, YXiang, WSun, NWang, QGao, Pan, SZeng, MOsada, DLi2}. Despite fundamental differences, e.g., the parent state of the cuprates is an antiferromagnetic insulator whereas the parent compounds of the nickelates are metals, the two systems share multiple clear similarities. Most notably, both have a broad superconducting dome that is flanked by Fermi-liquid behavior on the overdoped side and poor conductivity (a correlation-driven resistivity upturn in the nickelates) on the underdoped side~\cite{Lee}. 

A hallmark of cuprate superconductors is the evidence for vortex-like fluctuations of the superconducting order parameter that extend to temperatures $T$ well above the resistive critical temperature $T_{\rm c}$~\cite{Xu}. 
Above $T_{\rm c}$, the vortex liquid state persists to temperatures as high as 100 K above $T_{\rm c}$ in underdoped La$_{2-x}$Sr$_x$CuO$_4$ (LSCO) and YBa$_2$Cu$_3$O$_{6+y}$ (YBCO). This unusual behavior was discovered through Nernst~\cite{Xu, Wang1, Wang2, Wang3, Wang4} and torque magnetometry measurements of the diamagnetic response~\cite{Wang5, LLi1, LLi2, Johnston, LLi5,Yu}, which can detect the vortex-liquid state with high sensitivity. The rapid diffusion of the vortices leads to loss of phase rigidity in the condensate above $T_{\rm c}$. 

To sharpen the comparison between the two systems, we have investigated the vortex-Nernst effect in Nd$_{1-x}$Sr$_x$NiO$_2$ near $T_{\rm c}$ (6.5 - 15 K). We have also measured the low-temperature Seebeck effect which probes the particle-hole asymmetry in the normal-state density of states ${\cal N}(E)$. In a thermal gradient $-\nabla T\parallel {\bf\hat{x}}$ and a magnetic field ${\bf H\parallel \hat{z}}$, the observed electric field $\bf E$ in the $x$-$y$ plane defines the thermoelectric matrix $S_{ij}$, viz. $E_i = S_{ij}(-\partial_j T)$. The diagonal component $S_{xx}$ corresponds to the Seebeck effect (or thermopower) whereas the off-diagonal component yields the Nernst response
\be
S_{yx}(T, H) = E_y/|{\bf\nabla}T|.
\label{eq:Nernst}
\ee

Here the observed field profile of $S_{yx}$ is strongly non-linear in $H$, displaying a broad maximum at $\mu_0H \approx$ 5 T and a long tail that extends to 15 T. (Although our maximum field is 13 T, we estimate by extrapolation that the vortex signal vanishes at $\sim$15 T.) 
In an optimally doped sample ($x$ = 17.5\%), $S_{yx}$(5 T) falls from 290 nV/K at 6.5 K to less than 20 nV/K (below the noise floor of our experiment) at 15 K, concurrently with the onset of the normal-state resistance. However, $H_{c2}(T)$ does not vanish at $T_{\rm c}$. A similar trend occurs in a second sample with 20\% hole doping. From these results, we infer that vortex-like fluctuations exist in a temperature interval above $T_{\rm c}$, but the interval is smaller than in underdoped cuprates. We caution, however, that our results are taken on samples with known structural disorder and inhomogeneity~\cite{DLi1, Lee} which could substantially reduce the extent of the vortex liquid. It is possible that the Nernst response may extend to higher temperatures in improved films, such as those grown on (LaAlO$_3$)$_{0.3}$(Sr$_2$TaAlO$_6$)$_{0.7}$ substrates~\cite{Lee}, which warrants further investigation.

\section{Methods}
\begin{figure*}
    \centering
  \includegraphics[width=\textwidth]{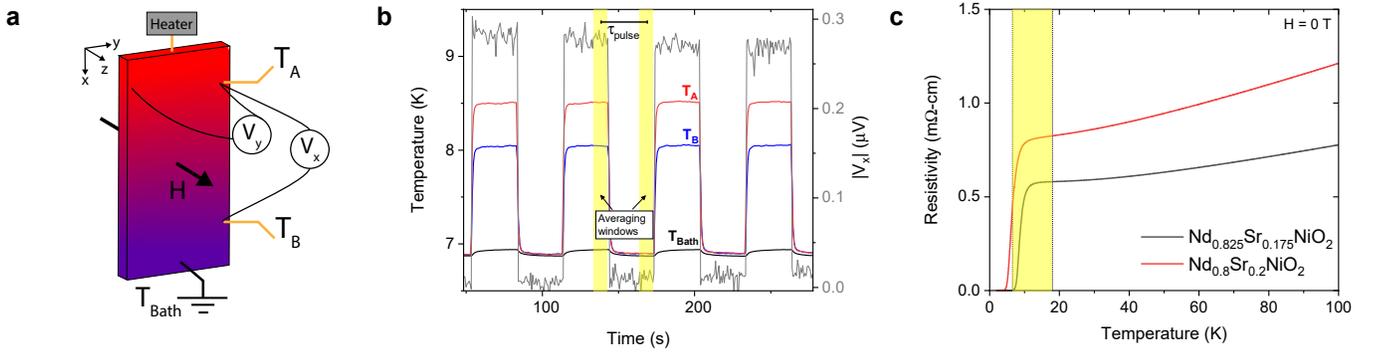}
    \caption{{\bf Experimental set-up. a.} Schematic of the Seebeck/Nernst experiment. A resistive heater fixed to the top end of the sample generates a temperature gradient $-\nabla T\parallel {\bf\hat{x}}$ that is measured by two RuO$_2$ thermometers $T_{A, B}$. The Seebeck response is calculated from the resulting longitudinal DC voltage $V_x$ and the Nernst from $V_y$. The magnetic field $H$ is directed perpendicular to the $a$-$b$ plane ($\parallel{\bf\hat{z}}$). {\bf b.} DC offsets and drifts are removed with a pulsed measurement technique. The heater is energized for a time $\tau_{\rm pulse}$, longer than the equilibration time between the sample and the temperature bath $T_{\rm bath}$, about 40 seconds. The voltage is recorded as the difference in the stabilized signals (yellow averaging windows) of the off/on stages of the cycle. {\bf c.} Zero-field resistivity versus temperature of the two nickelate films. Sample A (Sr = 17.5\%, optimally doped) is shown in black and Sample B (Sr = 20\%) in red. The yellow window highlights the temperature range accessible to our Nernst/Seebeck experiments. 
    }
    \label{fig:Setup}
\end{figure*}

Thin Nd$_{1-x}$Sr$_x$NiO$_2$ films (of thickness 9-11 nm) were grown on 2.5 $\times$ 5 mm$^2$ SrTiO$_3$ substrates by pulsed-laser deposition and capped with a 2 nm-thick SrTiO$_3$ top layer. (See Ref.~\cite{DLi1} for thin-film growth details.) We consider two samples with Sr concentration $x$ = 17.5\% (Sample A, optimally doped) and 20\% (Sample B, overdoped). In Sample A, the resistive transition begins at 15 K and reaches zero resistance at 6.1 K. The resistance is 10\% of its normal-state value at $T_{c,10\%R}$ = 7.7 K. The $T_{\rm c}$ in Sample B is lower, occurring between 3 and 13 K; $T_{c,10\%R}$ = 5.3 K. Under an applied temperature gradient, the base temperature reachable in our experiment is $\sim$6.5 K---slightly above the zero-resistance temperature of each sample  (Fig. \ref{fig:Setup}c).

Electrical characterization measurements were conducted with standard lock-in techniques using aluminum wire-bonded contacts. The thermoelectric DC voltages were probed by phosphor-bronze wires connected with silver paint to the aluminium bonds and soldered to pads on the cold-finger of a custom high-vacuum probe with all-copper wiring to room temperature. The voltages were read out by a Keithley 2182a nanovoltmeter. The temperatures were measured by two RuO$_2$ thermometers attached to the sample by thick gold wires using silver epoxy. These thermometers were calibrated \textit{in situ} in zero magnetic field using a Cernox CX-1010 thermometer which has very low magnetoresistance in this temperature range. The magnetic field was swept at $\leq$ 0.2 T/min.

A temperature gradient $-\nabla T\parallel {\bf \hat{x}}$ was generated in the sample using a 1 k$\Omega$ thin-film resistor mounted to one end of the crystal with silver paint. The opposite end of the sample was glued to a brass stage on the cold-finger using GE varnish (Fig. \ref{fig:Setup}a). A major challenge in the experiment is posed by the feeble Nernst response (maximum Nernst voltage $V_{\rm N}\sim 300$ nV). The large thermal conductance of the SrTiO$_3$ substrate severely limits the temperature gradient, especially above 15 K. The use of sensitive thermometers as well as a pulsed measurement technique to remove offsets and drift in the DC voltages allowed us to overcome some of these difficulties (Fig. \ref{fig:Setup}b). A current pulse lasting $\tau$ $\approx$ 40 s was applied to the heater to establish a stable $-\nabla T$. The longitudinal (Seebeck) and transverse (Nernst) voltage drops were measured after stabilization. The heater was then turned off and the sample was allowed to relax for time $\tau$. The DC voltages were again measured and these were subtracted from the previous values in order to record the true thermoelectric voltage responses. The sample temperature and temperature gradient were recorded as the average of and difference between the two thermometers during the energized part of the cycle.

\section{Results and Discussion}
\begin{figure*}
    \centering
  \includegraphics[width=\textwidth]{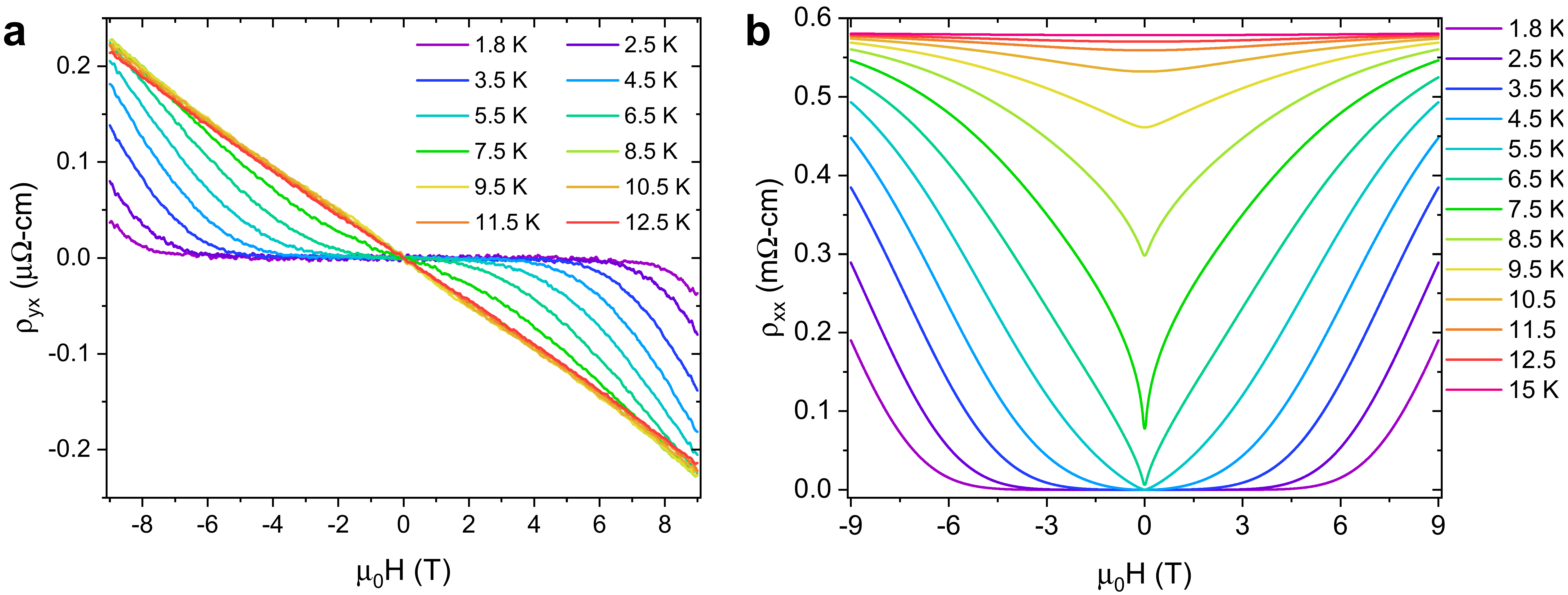}
    \caption{{\bf Magnetotransport ($ {\bf H}$ $\parallel$ ${\bf \hat{z}}$).} Hall effect ({\bf a}) and magnetoresistance (MR) ({\bf b}) in the optimally doped film, Sample A (Sr = 17.5$\%$). The sample reaches zero resistance at $T_{\rm c} \approx$ 6.1 K. We extract the vortex-solid melting field, $H_m(T)$ from the onset of non-zero MR ({\bf b}) and the onset field of the vortex-Hall effect $H_{\rm v}(T)$ from the appearance of non-zero $\rho_{yx}(H,T)$ ({\bf a}). These onset fields are plotted in Fig. \ref{fig:PhaseDiagram}c.
    }
    \label{fig:MRHall}
\end{figure*}

Figure \ref{fig:MRHall} depicts low-temperature magnetotransport measurements on the optimally doped film (Sample A, $x = 17.5\%$). In the normal state ($T > 13$ K), the Hall coefficient is $-2.7\times 10^{-4}$ cm$^2$/C, in agreement with previous measurements~\cite{DLi2}. Below $T = 6.5$ K, both the transverse (Hall) resistivity $\rho_{yx}$ and longitudinal resistivity $\rho_{xx}$ are zero up to an onset magnetic field: $H_m(T)$ in $\rho_{xx}$ and $H_{\rm v}(T)$ in $\rho_{yx}$. Magnetic fields stronger than $H_m(T)$ induce melting of the vortex solid. Rapid motion of vortices in the liquid state produce a finite $\bf E= B\times v$ observed as a finite flux-flow resistivity $\rho_{xx}$ and flux-flow Hall resistivity $\rho_{yx}$ ($\bf B$ is the flux density and $\bf v$ the velocity of the vortex core). We plot these onset magnetic fields in the $T$-$H$ phase diagram depicted in Figure \ref{fig:PhaseDiagram}c.

We have measured the thermoelectric matrix $S_{ij}$ in Nd$_{1-x}$Sr$_{x}$NiO$_2$ in the vicinity of the superconducting regime. In the normal state, the zero-$H$ Seebeck coefficient $S_{xx}(T) = E_x/|\nabla T|$ is proportional to $(d{\cal N}/dE)_\mu$, the derivative of the density of states ${\cal N}(E)$ evaluated at the chemical potential $\mu$. The Nernst response is comprised of two distinct contributions. The quasiparticle (or carrier) contribution $S^{\rm qp}_{yx}$ is distinguished by its $H$-linear field profile at moderate $H$ whereas the vortex-Nernst term $S^{\rm v}_{yx}$ has a characteristic non-monotonic field profile. By additivity, we have 
\be
S_{yx} = S^{\rm qp}_{yx} + S^{\rm v}_{yx}.
\label{addS}
\ee
In this system, the short quasiparticle mean-free-path $\ell$ makes $S^{\rm qp}_{yx}\ll S^{\rm v}_{yx}$ in general~\cite{Wang1,Wang4}. However, when the vortex-Nernst contribution vanishes (at $T>13$ K), $S^{\rm qp}_{yx}$ is readily resolved. When Cooper-pair correlations become large, phase slippage induced by moving vortices produces a dominant vortex-Nernst contribution $S^{\rm v}_{yx}$ which displays the characteristic tilted-dome field profile~\cite{Wang4} shown in Fig. \ref{fig:SyxSampleA}a. 

\begin{figure*}
    \centering
  \includegraphics[width=\textwidth]{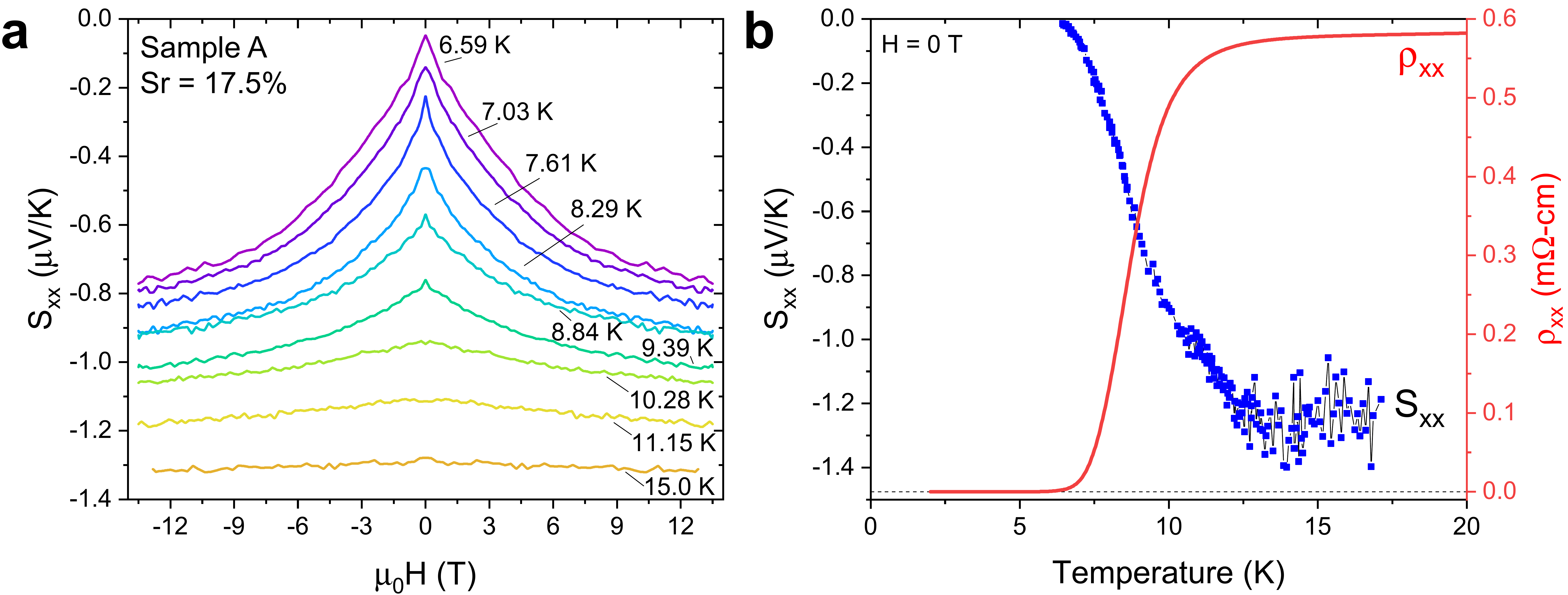}
    \caption{{\bf Seebeck effect in Nd$_{0.825}$Sr$
_{0.175}$NiO$_2$. a.} Magnetic field sweeps of the Seebeck coefficient $S_{xx}$($H, T$) at fixed temperature. {\bf b.} $S_{xx}(0, T)$ plotted alongside the zero-field resistivity ($\rho_{xx}$). $S_{xx}$ reaches zero concurrently with the onset of the zero-resistivity state at $T_{\rm c}$.}
    \label{fig:SxxSampleA}
\end{figure*}

\begin{figure*}
    \centering
  \includegraphics[width=\textwidth]{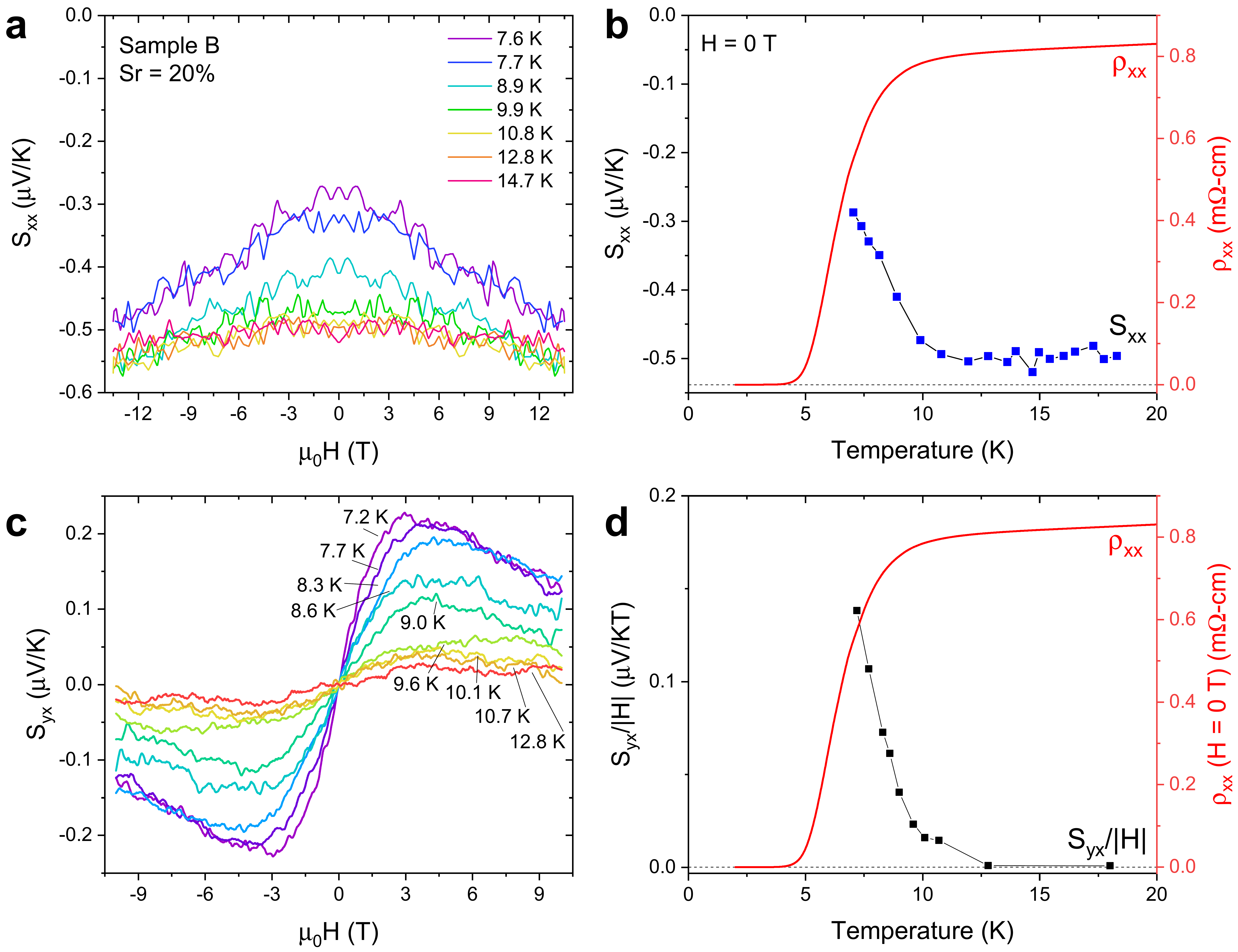}
    \caption{{\bf Seebeck and Nernst effects in Nd$_{0.8}$Sr$
_{0.2}$NiO$_2$ (Overdoped). a.} Field sweeps of the Seebeck coefficient $S_{xx}$($H, T$) at fixed temperature. {\bf b.} Temperature dependence of $S_{xx}$($H$ = 0, $T$), plotted alongside the zero-field resistivity ($\rho_{xx}$). In the normal state, $T$ = 12 - 18 K, $S_{xx}(H$ = 0, $T$) has a constant value of -0.5 \textmu V/K. {\bf c.} Field sweeps of the Nernst response $S_{yx}$($H, T$) at fixed temperatures. {\bf d.} The temperature dependence of the Nernst coefficient $\nu = S_{yx}/H$, plotted with the zero-field resistivity ($\rho_{xx}$). $S_{yx}/H$ vanishes by $T\approx$ 13 K.}
    \label{fig:SampleB}
\end{figure*}

Figure \ref{fig:SxxSampleA} shows the results of the Seebeck experiment on Sample A (Figs. \ref{fig:SampleB}a, b for Sample B). In the normal state, $T \geq 13$ K, the zero-field Seebeck coefficient $S_{xx}(T,0)$ is a negative constant: -1.2 \textmu V/K (-0.5 \textmu V/K) in Sample A (B). The negative sign of $S_{xx}$ agrees with that of the Hall coefficient (Fig. \ref{fig:MRHall}a). As $T$ is lowered below 15 K, $|S_{xx}|$ decreases monotonically, approaching zero at 6.5 K. The application of a perpendicular magnetic field $H$ suppresses superconductivity and increases $|S_{xx}|$ (Sample A: Fig. \ref{fig:SxxSampleA}a; Sample B: Fig. \ref{fig:SampleB}a). 

To address the vortex-Nernst response in the vicinity of $T_{\rm c}$, we distinguish between two types of fluctuations of the complex superconducting order parameter $\hat\Psi$. Gaussian (or amplitude) fluctuations correspond to fluctuations of the modulus $|\hat\Psi|$ about the minimum of the Ginzburg-Landau potential well. In a conventional type-I superconductor (Al and Sn), Gaussian fluctuations are restricted to a very narrow interval above $T_{\rm c}$ (of width $\delta T\sim 10^{-2}$ $T_{\rm c}$). 

The second class of fluctuations (called ``singular'') are produced by mobile vortices in the vortex-liquid state in type-II superconductors. In a driving temperature gradient $-\nabla T$, the passage of each vortex across the line joining the transverse ``Hall'' contacts causes the Josephson phase $\varphi$ between the contacts to wind by 2$\pi$~\cite{Xu,Wang1,Wang2,Wang3,Wang4}. For a vortex flow rate $\dot{N}_{\rm v}$ in the low-density limit (where $N_{\rm v}$ is the number of vortices), the Hall contacts detect the Josephson voltage $V_{\rm J}= 2\pi(\hbar/e) \dot{N}_{\rm v}$, which accounts for the vortex-Nernst signal $V_{\rm N}$ in weak $H$ (vortex cores are well-separated). The slope of the initial $H$-linear increase defines the Nernst coefficient, viz. 
\be 
\nu(T)= {\rm lim}_{H\to 0}\; S_{yx}(T,H)/H.
\label{nu}
\ee
When $S^{\rm v}_{yx}$ is dominant, the initial $H$-linear dependence reflects the insertion of vortices by the external field which leads to a proportionate increase of $V_{\rm N}$ with net vorticity. With increasing $H$, however, the vortex-Nernst response increasingly deviates downwards from the initial $H$-linear growth because of the decrease in $|\mathbf{v}|$ in large $H$ and the progressive weakening of the condensate amplitude $|\hat\Psi|$. Eventually, as $H\to H_{c2}$, $V_{\rm N}$ approaches zero~\cite{Wang4}. 

In conventional 3D type-II superconductors (e.g. Nb and NbSe$_2$), the vortex liquid in the $T$-$H$ plane is confined between the vortex-solid melting curve $H_{\rm m}(T)$ and the upper critical field $H_{c2}(T)$ (Fig. \ref{fig:PhaseDiagram}a). In the BCS scenario $H_{c2}(T)$ vanishes linearly with the reduced temperature $t = 1- T/T_{\rm c}$ as we approach $T_{\rm c}$ from below. Hence, in the $T$-$H$ plane, the two curves $H_{\rm m}(T)$ and $H_{c2}(T)$ converge to the point $(T_{\rm c},0)$ as $t\to 0$. Above $T_{\rm c}$, the vortex signal vanishes rapidly as a weak, evanescent tail in accord with the vanishing of $|\hat\Psi|$ at $T_{\rm c}$.

Fluctuations in hole-doped cuprate superconductors are categorically different from the BCS (mean-field) scenario~\cite{Xu,Wang4,Wang5,LLi5,Yu}. Below $T_{\rm c}$ (as measured by resistivity $\rho$) the vortex liquid remains bounded from below by the melting curve $H_{\rm m}(T)$ (Fig. \ref{fig:PhaseDiagram}b). However, the upper boundary $H_{c2}(T)$ has eluded accurate measurement (with estimates varying from 100 T to 300 T). A striking feature is that, as we approach $T_{\rm c}$ from below, we never observe the curve $H_{c2}(T)$ approaching zero in \emph{any} of the hole-doped cuprates~\cite{Wang4,LLi5}. Instead, the vortex-liquid state is observed to extend well above $T_{\rm c}$ to temperatures as high as 160 K. These unexpected observations have led to the conclusion that, in the hole cuprates, the amplitude $|\hat\Psi|$ remains finite well above $T_{\rm c}$~\cite{Wang4,Wang5,LLi5,Yu}. The abrupt onset of finite  $\rho$ at $T_{\rm c}$ actually corresponds to the vanishing of phase rigidity and the emergence of the vortex-liquid state; the Cooper condensate remains stable above $T_{\rm c}$ but is strongly phase disordered. Aside from the Nernst evidence, this scenario is confirmed by measurements of the diamagnetic response using torque magnetometry~\cite{Wang5,LLi1,LLi2,LLi5,Yu}. A closely related observation is that the superconducting energy gap $\Delta$ measured in tunneling experiments persists at the same value to temperatures well above $T_{\rm c}$ (instead of closing) while the subgap region is progressively filled in. With this perspective in mind, we examine the situation in the infinite-layer nickelate. 

The key result in our report is Figure \ref{fig:SyxSampleA}a which shows the field profiles of the observed Nernst response $S_{yx} = E_y/|{\bf\nabla} T|$ at various temperatures $T$ above $T_{\rm c}$ in the optimally doped sample (Sample A). Most of the curves feature the characteristic tilted-dome profile~\cite{Wang4}. As $H$ increases, the initial $H$-linear form in weak $H$ evolves to a broad maximum near 5 T, finally followed by a gradual monotonic decrease on the large-$H$ side. As discussed, the field at which the tilted-dome profile ends nominally indicates where the condensate amplitude vanishes~\cite{Wang4,Wang5,LLi5,Yu}. This provides an estimate of the upper critical field $H_{c2}$.

However, if the quasiparticle contribution $S^{\rm qp}_{yx}$ is finite (see Eq. \ref{addS}), we need to subtract off $S^{\rm qp}_{yx}$ before estimating $H_{c2}$~\cite{Wang1}. The $H$-linear field profile of $S_{yx}(T,H)$ measured at 15.03 K (lowest-lying curve in Fig. \ref{fig:SyxSampleA}a) suggests that it is predominantly comprised of $S^{\rm qp}_{yx}$. 

This is confirmed by examining the $T$ dependence of the Nernst coefficient $\nu$ (Eq. \ref{nu}) plotted in Fig. \ref{fig:SyxSampleA}c. At $T = 6.5$ K (just above $T_{\rm c}$), $\nu$ is observed to have the rather large value 180 nV/KT. As $T$ is raised above $T_{\rm c}$, $\nu$ decreases very steeply but saturates (at $\sim 13$ K) to a small finite value ($\sim 10$ nV/KT). The observed behavior supports our identification of the curve of $S_{yx}$ at 15.03 K in Fig. \ref{fig:SyxSampleA}a with the quasiparticle contribution $S^{\rm qp}_{yx}$. The onset temperature of the vortex Nernst signal is estimated as $T_{\rm onset}\simeq 13$ K.

Hence, subtracting the curve of $S_{yx}(H)$ at 15.03 K from the other curves at lower $T$ in Fig. \ref{fig:SyxSampleA}a isolates the field profile of the vortex Nernst signal. At a selected temperature $T'$, we estimate $H_{c2}(T')$ as the field at which the curves of $S_{yx}$ measured at $T'$ and at 15.03 K intersect.

Similar features are obtained in Sample B (Fig. \ref{fig:SampleB}c, d).

We note that estimates of the upper critical field based on the resistive transition (which we call $H^{\rm R}_{c2}$ here) show that $H^{\rm R}_{c2}(T)$ approaches zero as $T\to T_c$~\cite{BYWang1, BYWang2} instead of the $T$-independent profile in Fig. \ref{fig:PhaseDiagram}c. The resistive transition actually measures the collapse of phase rigidity caused by melting of the vortex solid, which leads to dissipation, whereas the vortex Nernst signal detects the existence of the pair condensate. In a mean-field superconductor, both should agree. However, when phase rigidity vanishes in relatively weak $H$, they strongly disagree. This is the situation in hole-doped cuprates and appears to be the case here as well.

\section{Vortex-liquid state}
From the results in Sample A, we derive the phase diagram shown in Figure \ref{fig:PhaseDiagram}c. The onset field of dissipation measured by the resistivity $\rho_{xx}(T,H)$, which we define to be the melting field $H_m(T)$ is plotted as black squares. The onset field $H_{\rm v}$ of the vortex Hall effect $\rho_{yx}(T,H)$ is plotted as blue circles. At each $T$, the relation $H_{\rm v}(T)>H_m(T)$ implies that melting of the vortex solid initially leads to the diffusive motion of vortices that cause $\rho_{xx}$ to be finite. However, the core velocity $\bf v$ becomes large enough to produce a finite $\rho_{yx}$ only when $H$ exceeds the higher $H_v(T)$. 
The relation $H_{\rm v}(T)>H_m(T)$ was previously observed in $2H$-NbSe$_2$~\cite{Jing}. For comparison, we also plot the profile of the zero-$H$ resistivity $\rho$ vs. $T$ (dashed curve). 

\begin{figure*}
    \centering
  \includegraphics[width=0.8\textwidth]{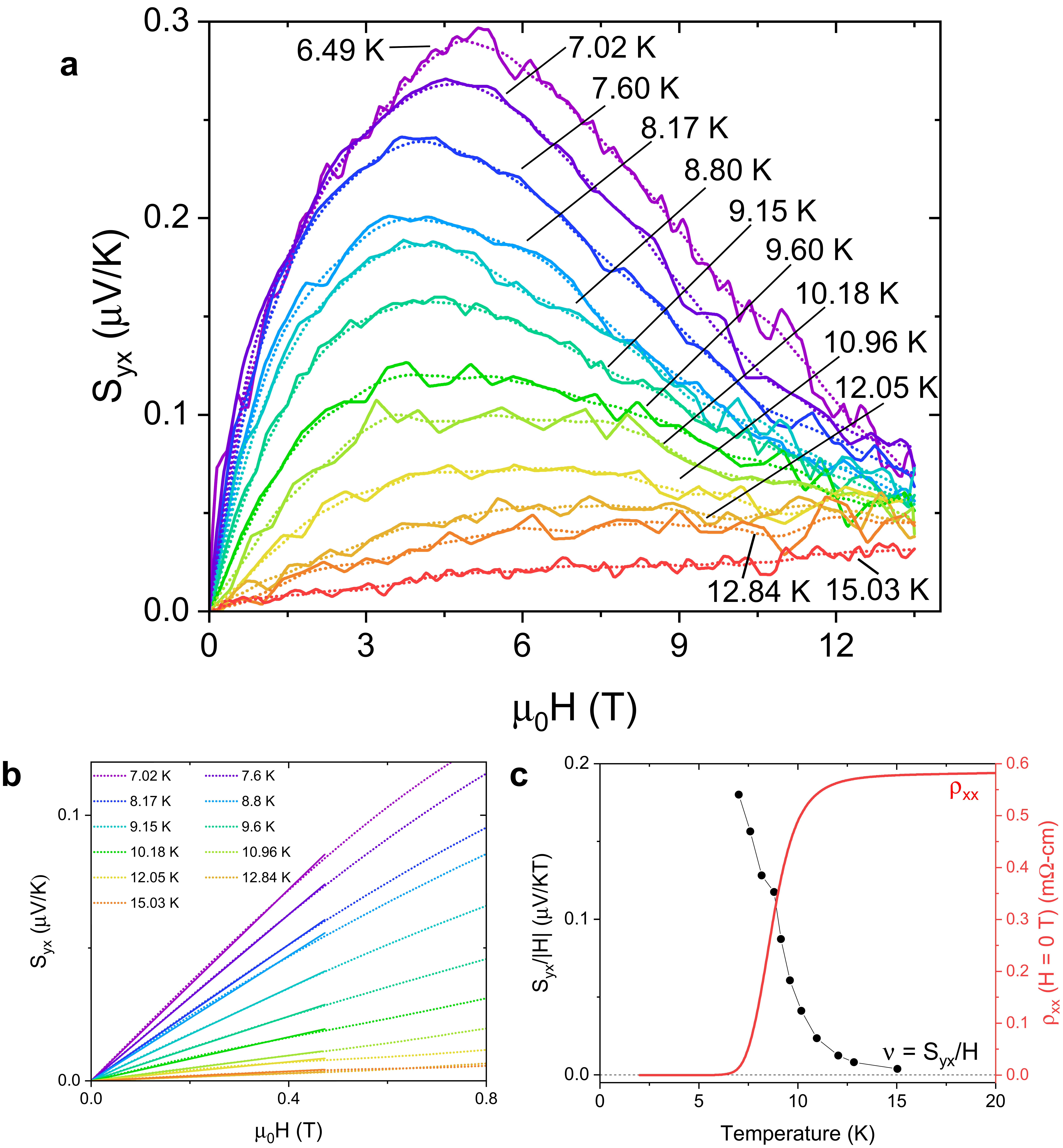}
    \caption{{\bf Vortex-Nernst effect in Nd$_{0.825}$Sr$
_{0.175}$NiO$_2$.} {\bf a.} Field sweeps of the Nernst response $S_{yx}$($H, T$) at fixed temperatures. The solid-line curves are the field-antisymmetrized raw data and the dotted lines have Savitsky-Golay smoothing added. {\bf b.} The low-field ($\leq$ 0.5 T) linear fits used to extract $\nu$ in {\bf b}. The dotted lines are the smoothed $S_{yx}$ data shown in  a. {\bf c.} This  shows the temperature dependence of the Nernst coefficient $\nu = S_{yx}/H$ alongside the zero-field resistivity. The Nernst coefficient falls below 10 nV/KT at $T >$ 13 K.}
    \label{fig:SyxSampleA}
\end{figure*}

\begin{figure*}
    \centering
  \includegraphics[width=0.7\textwidth]{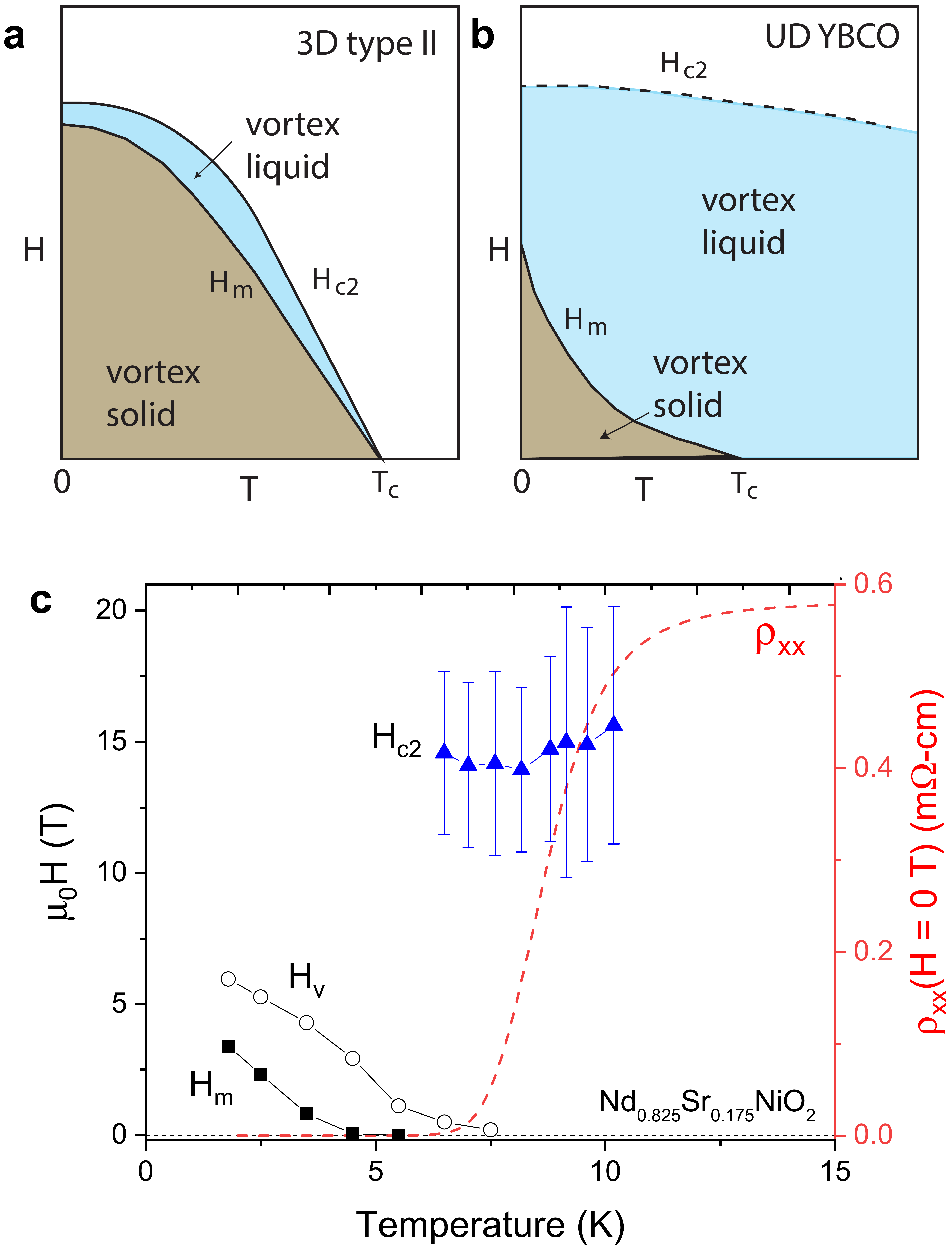}
    \caption{{\bf Comparison of the BCS, underdoped cuprate, and infinite-layer nickelate $T$-$H$ phase diagrams. a.} The standard BCS phase diagram in the $T$-$H$ plane for a 3D type-II superconductor. The vortex solid (grey) is stable below the melting field $H_m(T)$. The vortex liquid (blue) is confined between $H_m(T)$ and the upper critical field $H_{c2}(T)$. $H_{c2}(T)$ decreases to zero linearly in $(T_c-T)$. {\bf b.} The phase diagram in an underdoped cuprate (e.g. YBa$_2$Cu$_3$O$_y$ ($y\sim$ 6.6)~\cite{Yu}). The melting field $H_m(T)$ shows pronounced positive curvature vs. $T$ and terminates at $T_{\rm c}$. The upper critical field $H_{c2}(T)$ (dashed curve) is estimated to lie well above 50 T. At $T_{\rm c}$, $H_{c2}$ remains very large. {\bf c.} The $T$-$H$ phase diagram for the infinite-layer nickelate inferred from the vortex-Nernst effect in Sample A. The lowest curve (solid squares) plots the onset of dissipative behavior in $\rho_{xx}(H)$ which we identify with $H_m(T)$. The field at which $\rho_{yx}$ becomes finite ($H_{\rm v}(T)$) is shown as open circles. The upper critical fields $H_{c2}(T)$ inferred from high-field linear fits to $S_{yx}(T,H)$ (Fig. \ref{fig:SyxSampleA}a) are plotted as blue triangles. The error bars represent the propagation of one standard deviation in the linear fit of $S_{yx}(T$ = 15 K$,H >$ 9 T$)$ subtracted from that of $S_{yx}(T,H >$ 9 T$)$. We infer that the nickelate vortex-liquid state exists in the region bounded above by $H_{c2}(T)$ and below by $H_m(T)$. Note the poor signal-to-noise ratio in $S_{yx}$($H,T>$ 11 K) precludes us from estimating $H_{c2}(T>$ 11 K).}
\label{fig:PhaseDiagram}
\end{figure*}

The comparison implies that the zero-field dissipationless state that emerges at $T_{\rm c}$ (6.1 K) is readily suppressed in a weak $H$. The most instructive plot is $H_{c2}(T)$ (blue triangles) inferred from $S^{\rm v}_{yx}$ in Fig. \ref{fig:SyxSampleA}a. 
We find that $H_{c2}(T)$ remains at the same value ($\sim$15 T) even as $T$ exceeds $T_{\rm c}$. As mentioned, the vortex-Nernst contribution to the coefficient $\nu$ remains finite until 13 K (Fig. \ref{fig:SyxSampleA}c). Together, the three curves $H_{\rm m}(T)$, $\nu(T)$ and $H_{c2}(T)$ imply that the entire dissipative region shaded in blue is actually a vortex-liquid state that persists to
$T_{\rm onset}\simeq 13$ K, well above $T_{\rm c}$ (6.1 K). The ratio $T_{\rm onset}/T_c\simeq 2$ is comparable to values obtained in optimally and overdoped La$_{2-x}$Sr$_x$CuO$_4$. 
This implies that the transition at $T_{\rm c}$ also corresponds to a loss of phase rigidity of the pair condensate rather than closing of the gap $\Delta(T)$. Based on this conclusion, we anticipate that $\Delta(T)$ should not close at $T_{\rm c}$ but remain finite up to 13 K as a pseudogap whose subgap states are progressively occupied.

In the present samples, the resistive transition is quite broad. This suggests that the films may be comprised of domains with a broad distribution of condensate strengths as reflected by the individual $T_{\rm c}$'s. However, we do not expect a broad distribution to affect our argument supporting a vortex-liquid state based on the curves in  Fig. \ref{fig:SyxSampleA}a. A distribution of superconducting domains, each following the BCS scenario (Fig. \ref{fig:PhaseDiagram}a), will still display an $H_{c2}(T)$ curve that approaches zero at the highest $T_{\rm c}$, in conflict with our observation. 

Recently, optimally doped Nd$_{1-x}$Sr$_x$NiO$_2$ grown on LSAT substrates have been shown to have higher crystalline quality than the present films grown on SrTiO$_3$~\cite{Lee}. The resistive transitions are considerably sharper with the $T$-linear resistivity profile persisting nearly to $T_{\rm c}$. Nernst measurements on the improved samples should allow us to determine how $T_{\rm onset}$ varies with doping.

\newpage
\vspace{1cm}
\noindent
$^\S$Corresponding author email: \href{mailto:npo@princeton.edu}{npo@princeton.edu}\\

\noindent
{\bf Acknowledgement}
We thank Kyuho Lee for helpful discussions. We acknowledge support of the U.S. National Science Foundation through grant DMR 2011750. Work at Stanford/SLAC was supported by the Department of Energy, Office of Basic Energy Sciences, Division of Materials Sciences and Engineering, under contract no. DE-AC02-76SF00515.

\vspace{3mm}
\noindent
{\bf Author contributions}\\
NPQ, NPO and HYH conceived the idea for the experiment. Samples were grown and characterized by DL, BYW, and HYH. NPQ performed all the measurements and wrote the text with NPO and inputs from all authors. 

\vspace{3mm}
\noindent
{\bf Competing financial interests}\\
The authors declare no competing financial interests.

\vspace{3mm}
\noindent

\vspace{3mm}
\noindent
{\bf Correspondence and requests for materials}
should be addressed to N.P.Q. or N.P.O.

\nocite{*}

\end{document}